%% file: main.tex
\documentclass[conference]{IEEEtran}
\IEEEoverridecommandlockouts
\usepackage{cite}
\usepackage{amsmath,amssymb,amsfonts}
\usepackage{graphicx}
\usepackage{textcomp}
\usepackage{xcolor}
\usepackage{geometry}
\usepackage{multicol}
\usepackage{tikz-cd}
\usepackage{mathrsfs}
\usepackage{float}
\usepackage{pstricks}
\usepackage{enumitem}
\usepackage{multirow}
\usepackage{listings}
\usepackage{subfiles}
\usepackage{algorithmic}
\usepackage{booktabs}
\usepackage{setspace}
\usepackage{array}

\def\BibTeX{{\rm B\kern-.05em{\sc i\kern-.025em b}\kern-.08em
    T\kern-.1667em\lower.7ex\hbox{E}\kern-.125emX}}
\begin{document}

\title{Sketch and Scale \\
Geo-distributed tSNE and UMAP

\thanks{This research was supported in part by NSF CAREER grant 1652257, NSF grant 1934979, ONR Award N00014-18-1-2364 and the Lifelong Learning Machines program from DARPA/MTO.}
}

\author{

\IEEEauthorblockN{Viska Wei}
\IEEEauthorblockA{\textit{Johns Hopkins University}\\
swei20@jhu.edu}
\and
\IEEEauthorblockN{ Nikita Ivkin}
\IEEEauthorblockA{\textit{Amazon}\\
ivkin@amazon.com}
\and
\IEEEauthorblockN{ Vladimir Braverman}
\IEEEauthorblockA{\textit{Johns Hopkins University} \\
vova@cs.jhu.edu}
\and
\IEEEauthorblockN{Alexander S. Szalay}
\IEEEauthorblockA{\textit{Johns Hopkins University} \\
szalay@jhu.edu}
}

\def \eps {\varepsilon}

\IEEEoverridecommandlockouts
\IEEEpubid{\makebox[\columnwidth]{978-1-7281-6251-5/20/\$31.00 ~\copyright2020 IEEE \hfill} \hspace{\columnsep}\makebox[\columnwidth]{ }}
\maketitle
\IEEEpubidadjcol
    \subfile{0-abstract}
    \subfile{1-intro}
    \subfile{2-countsketch}
    \subfile{3A-cancer}
    \subfile{3B-sdss}

    \subfile{4-results}


\bibliographystyle{acm}
\bibliography{ref}


\end{document}

%% file: 0-abstract.tex
\begin{abstract}
Running machine learning analytics over geographically distributed datasets is a rapidly arising problem in the world of data management policies ensuring privacy and data security. Visualizing high dimensional data using tools such as t-distributed Stochastic Neighbor Embedding (tSNE) and Uniform Manifold Approximation and Projection (UMAP) became a common practice for data scientists. Both tools scale poorly in time and memory.
While recent optimizations showed successful handling of 10,000 data points, scaling beyond million points is still challenging. We introduce a novel framework: Sketch and Scale (SnS). 
It leverages a Count Sketch data structure to compress the data on the edge nodes, aggregates the reduced size sketches on the master node, and runs vanilla tSNE or UMAP on the summary, representing the densest areas, extracted from the aggregated sketch. 

We show this technique to be fully parallel, scale linearly in time, logarithmically in memory and  communication, making it possible to analyze datasets with many millions, potentially billions of data points, spread across several data centers around the globe. 
We demonstrate the power of our method on two mid-size datasets: cancer data with 52 million 35-band pixels from multiplex images of tumor biopsies; and astrophysics data of 100 million stars  with  multi-color photometry from the Sloan Digital Sky Survey (SDSS).  

\end{abstract}

%% file: 1-intro.tex
\section{Introduction}
\vspace{-0.1cm}
Dimensionality reduction plays a crucial role in both machine learning and data science.
It primarily serves two fundamental roles: (1) as a pre-processing step it helps to extract the most important low dimensional representation of a signal before feeding the data into a Machine Learning algorithm, (2) as a visualization tool it navigates data scientists towards better understanding of local and global structures within the dataset 
while working with more comprehensible two- or three-dimensional plots.  In clustering and classification problems we often seek to find a relatively small number of clusters, 
to correspond the number of categories human perception can distinguish.

Among the full spectrum of dimensionality reduction and lower dimensional embedding techniques available today \cite{ rokhlin2010randomized, roweis2000nonlinear,sammon1969nonlinear, tenenbaum2000global, hinton2003stochastic}, tSNE \cite{maaten2008visualizing} and UMAP \cite{mcinnes2018umap} are probably the two most popular methods for visualization. t-distributed stochastic neighbor embedding (tSNE) is a high-dimensional data visualization tool proposed by Geoffrey Hinton's group in 2008 \cite{maaten2008visualizing}. tSNE converts similarities between data points to joint probabilities and tries to minimize the Kullback-Leibler divergence between the joint probabilities of the low-dimensional embedding and the high-dimensional data. In contrast to PCA, tSNE is not linear, it  employs the local relationships between points to create a low-dimensional mapping, by comparing the full high-dimensional distance to the one in the projection subspace and capturing non-linear structures. tSNE has a non-convex cost function and provides different results for different initialisations.

One of the major hurdles with tSNE is that it ceases to be efficient when processing high- (or even medium) dimensional data sets with large cardinalities, due to the fact that the naive tSNE implementation scales as $O(n^2)$. On a typical laptop CPU, processing 10,000 points with tSNE would take close to an hour. There was a considerable research effort exerted to speed up tSNE. Barnes-Hut approximation pushed scaling down to $O(n \log n)$ \cite{van2014accelerating}. 
Approximated tSNE \cite{pezzotti2016approximated} lets user steer the trade off between speed and accuracy; netSNE is training neural networks on tSNE \cite{cho2018neural} to provide a more scalable execution time. The multicore tSNE~\cite{Ulyanov2016} and tSNE-CUDA~\cite{chan2018t} introduced highly parallel version of the algorithm for CPU and GPU platforms.
UMAP \cite{mcinnes2018umap} is using manifold learning and topological data analysis to reduce dimensionality. It uses cross-entropy to optimize the lower dimensional representation. Faster performance is the main advantage over tSNE~\cite{umapbench}. 
Nevertheless, both frameworks scale poorly and computational prohibitive when data hits 
a hundred million points.
In addition, the entire dataset has to reside in memory of one machine, making it infeasible for the datasets distributed across several compute clusters around the globe. For instance, privacy concerns of the healthcare data might limit transfers from clinic to clinic, physics related data accumulated by several research centers can be too large to transfer.   

In this paper, we introduce \textit{Sketch and Scale} (SnS), a solution to deal with much higher-cardinality datasets with an intermediate number of dimensions. 
We implement our idea as a preprocessor to any of the above mentioned dimensionality reduction techniques. SnS uses approximate sketching over parallel, linear streams. Furthermore, it can be executed over spatially segregated subsets of the data. Specifically, we utilize a hashing-based algorithm Count Sketch over the quantized high dimensional coordinates to find the cells with the highest densities: the so called ``heavy-hitters''~\cite{charikar2002finding,cormode2005improved}.
We then select an appropriate number of heavy hitters ($10^4$-$10^5$) to analyze with the standard techniques to get the final clusters/visualizations.

In Section~\ref{sec:scalable_preprocessor} we present the idea of sketching and describe its crucial role in building scalable pre-processing pipeline. Further, in Section~\ref{sec:app} we apply it to two data sets: multispectral cancer images, with a total of 52 million pixels, and photometric observations of 32 million stars from the Sloan Digital Sky Survey. Finally  we discuss various practical aspects on how to scale our technique to much larger data sets, geographically disjoint data and overcoming privacy-related constraints.

\section{Scalable preprocessor to tSNE/UMAP}\label{sec:scalable_preprocessor}

There is a genuine need to find clusters in data sets with cardinalities in the billions: pixels of multispectral imaging data in medical and geospatial imaging, large multi-dimensional point clouds, etc. Furthermore, we want to compute approximate statistics (various moments) of a multidimensional probability distribution with a large cardinality. Our tool helps with all of this: it generates a very compact approximation to the full multidimensional probability distribution of a large~dataset.

\subsubsection{\textbf{Clusters and heavy hitters}}
Hereafter we will assume that our data is clustered, i.e. there exists a metric space with a non-vanishing correlation function: the excess probability over random that we can find two points at a certain distance from one another. While a Gaussian random process can be fully described with a single correlation function, other higher order processes can have non-trivial higher order,  N-point correlations. We can quantify this by creating a discretized grid over our metric space, and counting the number of points in each bin. This way we introduce a probability distribution $P(N)$  that a bin will count $N$ points.  

\begin{figure}[t]
\centering
\includegraphics[width=0.7\linewidth
]{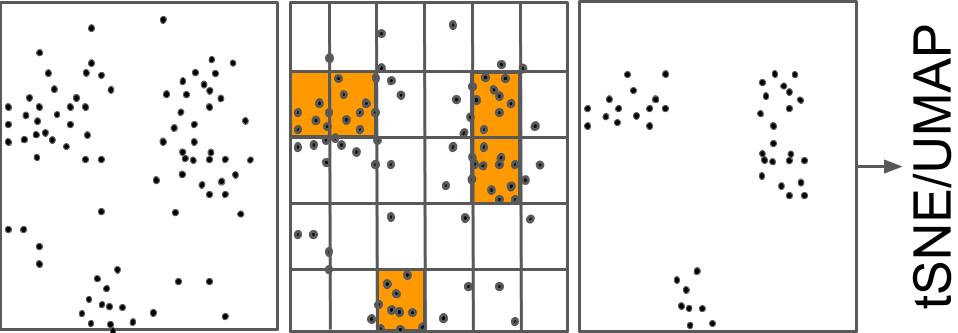}
\caption{ \footnotesize Preprocessing steps: 1. set a regular grid; 2. count points in every bin and find heavy bins; 3. create several represantatives for each heavy bin; 4. feed representatives into tSNE/UMAP.}
\label{fig:pipeline}
\vspace{-0.6cm}
\end{figure}

The bins intersecting our biggest clusters will have a large count. We will call these {\em ``heavy hitters''}~(HH). Datasets with strong clustering will have cell count distributions with a fat tail. Many of these heavy hitters will be contact neighbors, as the real clusters will be split by the grid discretization. A way to identify heavy hitters is to count the number of points in each bin, those above the threshold will be our heavy hitters. 
Finally, we use vanilla tSNE/UMAP to find the real, connected clusters which have been split up into multiple adjacent bins. tSNE/UMAP is applied in the reduced cardinality of the heavy hitters, each representing potentially millions of the original data points.  Refer to Fig.~\ref{fig:pipeline} for high level overview of the preprocessing pipeline.

We weighted each HH by replicating it multiple times with small uniform perturbation ($1/4$ of the cell size), since identical points are merged in tSNE. 
One scheme is to give a higher score towards the highest ranked HH. Assume that the smallest HH has a rank of $r_{max}$, and a count $f_{min}$. A second possibility for weighting is to use $1+\lfloor\log_2(r_{max}/r)\rfloor$ as the number of the replicas, scaling with the rank $r$ of the HH. Finally, we can also use $1+\lfloor\log_2(f/f_{min})\rfloor$ as the replication factor, weighting by the log of the counts $f$.  We have tested that these do not impact the main features of the clustering patterns we find. 

\subsubsection{\textbf{Heavy hitters vs random subsampling}}
One would think we apply tSNE/UMAP to the subsampled data, 
however, this does not work at low sampling rates. No matter how fat the tail of the original $P(N)$ distribution is, it will converge to an infinitesimally Poisson process as the sampling rate nears zero, and the fat tail rapidly disappears~\cite{szapudi1996szalay}.

Consider a data set with $10^{11}$ points with some dense clusters of $10^7$ points or more. To apply tools like tSNE/UMAP, we need to reduce the dataset size down to around $10^4$, i.e. sampling rate of $10^{-7}$. In order to detect a cluster with even a modest significance, we need to have enough points to keep the Poisson error low. In practice 100 points or more will push relative Poisson error to $1/\sqrt{100}=0.1$. For instance, a dense cell with $N\approx 10^7$ points, sampled at a rate of $10^{-7}$, would have in average only one point sampled, $K=N p = 1$. Therefore, it will be indistinguishable from the many billions of low density cells. 
At a sampling rate of $p=10^{-5}$ we could detect $K=100$ for the cell with $10^6$ points with a 10\% uncertainty, but the whole random subset would have $10^{11-5}=10^6$ points, too large to feed it into tSNE. With a larger data set these tradeoffs get rapidly much worse. 


In contrast, we detect the top heavy hitters with a high confidence, using direct aggregation at full or somewhat reduced sampling rate.  Then we take a small enough subset of these, so that tSNE/UMAP can easily cluster them further, while we discard all the low-density bins. This has the advantage, that a small number of the top heavy hitters will still contain a large fraction of the points, particularly those which are located in dense clusters. Of course, aggregating a multidimensional histogram in high resolution with a large number of dimensions by brute force would result in an untenable memory requirement. Instead we will use sketching techniques to build an approximate aggregation with logarithmic memory and linear compute time.

Our approach is particularly powerful for the data with the following properties: (i) the data has a very large number of rows ($10^8$ to $10^{12}$), (ii) has a moderate number of dimensions ($<20$) and (iii) there is a moderate number of clusters ($<100$), but these can have non-compact extents (iv) the clusters have a high density contrast in the metric space. The limitation on the number of dimensions is less of a problem than it seems, as dimensionality reduction techniques, like random projections~\cite{shah2019} can be applied as a preprocessor to our code.

\subsubsection{\textbf{Streaming sketches of heavy hitters}} \label{sec:streaming}

The field of streaming algorithms arose from the necessity to process 
massive data in sequential order while operating in a very low memory. 
First introduced for the estimation of  the frequency moments~\cite{alon1999space}, it was further developed for a wide spectrum of problems within linear algebra ~\cite{woodruff2014sketching}, graph problems~\cite{mcgregor2014graph}, and others~\cite{muthukrishnan2005data}; and found applications in machine learning\cite{ivkin2019communication, rothchild2020fetchsgd}, networking\cite{ivkin2019know,liu2016one},  astrophysics\cite{ivkin2018scalable, liu2015streaming}. Further we provide a glimpse on streaming  model and sketches
for finding frequent items. For comprehensive review refer to~\cite{muthukrishnan2005data}. 
Given a zero vector $f$ of dimension $n$, the algorithm observes the stream of updates to its coordinates $S = \{s_1, ..., s_m \}$, where $s_j$ specifies the update to $i$: $f_i \leftarrow f_i + 1$.  Alon, Matias and Szegedy~\cite{alon1999space} were first to show the data structure (AMS sketch) approximating $\ell_2$ norm of the vector $f$ at the end of the stream while using only $O(\log nm)$ bits of memory. In a nutshell, AMS is a counter $c$ and a hash function $h: [m] \rightarrow \{-1, +1\}$, on arrival of $s_j$, $c$ is updated as  $c = c + h(s_j)$. At the end of the stream, $c^2$ is returned as an approximation of $\|f\|_2^2$ . It is unbiased: 

\small
\vspace{-0.5cm}
$$E(c^2) = E(\sum_{j=1}^mh(s_j))^2 = E(\sum_{i=1}^n f_i h(i))^2 = $$
\vspace{-0.3cm}
$$ E\sum_{i=1}^n f_i^2h^2(i) + E\sum_{i\neq j} f_i f_j h(i)h(j)= \sum_{i=1}^n f_i^2 = \|f\|^2_2,$$

\normalsize
where the last inequality holds due to $h^2(i) = 1$ and $E(h(i)h(j)) = 0$ for $i\neq j$ and 2-wise independent~$h(\cdot)$.
Similarly, one can show the variance bound: $Var(c^2) \le 4 \|f\|^2_2$, then running several instances in parallel, averaging and/or using a median filter provide control over the approximation \cite{alon1999space}.

Count Sketch (CS) algorithm \cite{charikar2002finding} extends AMS approach to find heavy hitters with an $\ell_2$ guarantee, i.e. all $i$, s.t. $f_i > \eps \|f\|_2$, together with approximations of $f_i$. The idea behind the algorithm combines the AMS sketch with a hashing table with $C$ buckets. 
It uses hash $h_1(\cdot)$ to map arrived item $i$ to one of the $C$ buckets and hash $h_2(\cdot)$ for choosing the sign in AMS~sketch. Every  $(\eps, \ell_2)$-heavy hitter's  frequency can be estimated by the corresponding bin count if we choose  $C = 1/\eps^2$, as on average only $\eps$ fraction of $\ell_2$ norm of non-heavy items will fall into the same bin. To eliminate the collisions and identify the heavy hitters $R = \log(n/\delta)$ hash table maintained in parallel, where $1 - \delta$ probability of successful recovery of all heavy hitters. CS memory utilization is sublinear in $n$ and $m$:  $O(\frac{1}{\eps^2}\log nm)$, which is handy when working with billions and even trillions of items.
\begin{figure}[t]
\centering
\includegraphics[width=0.8\linewidth,height=0.07\textheight]{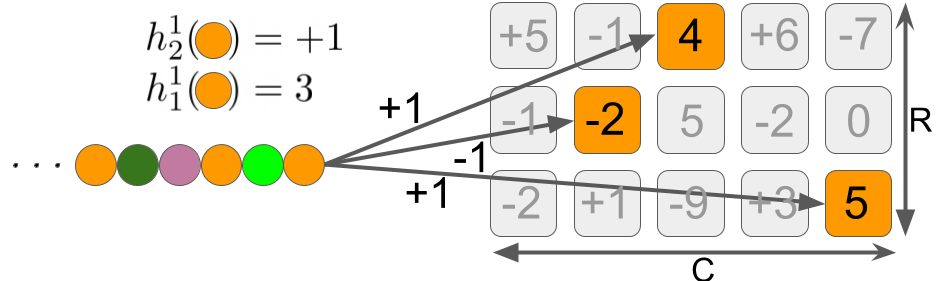}
\caption{\footnotesize Count Sketch update step: for each item compute two hashes for each row of the sketch, first hash chooses the bin, second hash chooses the sign of the change to apply (+1 or -1).}
\label{fig:input_data}
\vspace{-0.6cm}
\end{figure}

%% file: 2-countsketch.tex
\section{Scalable count sketch}\label{ap:cspipe}
\subsubsection{\textbf{The Count Sketch algorithm}}

As described in section~\ref{sec:streaming}, the Count Sketch (CS)  accepts updates to $n$-dimensional vector $f$ in a streaming fashion and can recover coordinates of $f$ for which $f_i \ge \eps \|f\|_2$, i.e. $(\eps, \ell_2)$-heavy hitters. It's major advantage is memory usage that scales logarithmically in dimensionality $n$ and stream size $m$. Moreover, CS is a linear operator, thus can be merged efficiently. This opens a diverse pool of applications in distributed settings: multiple nodes can compute sketches, each on its own piece of data, and send it to the master node for merging. Such an approach can help to alleviate the bottleneck in communication speed and brings a certain level of privacy for free~\cite{ivkin2019communication, rothchild2020fetchsgd}. Below we present the details of the four major operations over the CS data structure: initialization, update, estimate and merge \cite{charikar2002finding}.  

\vspace{1pt}
    \begin{algorithmic}[1]
    \label{code:cs}
        \footnotesize
        \STATE \textbf{function} init($R$, $C$):
        \STATE ~~~~ init $R\times C$ table of counters $S$ (with zeros)
        \STATE ~~~~ init bucket hashes: $\left\{ h_1^r: [n]\rightarrow [C]\right\}_{r=1}^R$
        \STATE ~~~~ init sign hashes: 
        $\left\{ h_2^r: [n]\rightarrow \pm 1\right\}_{r=1}^R$
        
        \STATE \textbf{function} update($s_i)$):
        \STATE ~~~~ \textbf{for} $r$ in $1\ldots R:~~$ $S[r, h_1^r(s_i)]$ += $h_2^r(s_i)$
    
        \STATE \textbf{function} estimate($i$):
        \STATE ~~~~ init length $R$ array estimates
        \STATE ~~~~ \textbf{for} $r$ in $1\ldots R$:~~ estimates$[r] = h_2^r(i) S[r, h_1^r(i)]$
        \STATE ~~~~ \textbf{return} median(estimates)
        
        \STATE \textbf{function} merge($S_1, S_2$):
        \STATE ~~~~ \textbf{return} $S_1 + S_2$

    \end{algorithmic}
\vspace{0pt}

\subsubsection{\textbf{Design of the experiment}}

We utilized GPU optimized Count Sketch\cite{Viska2020} running on CPU Intel Xenon Gold 6126 and GPU Tesla V100. CS parameters are set ~\cite{charikar2002finding} to find  $2\cdot10^4$ most frequent items from stream of length $10^8$: $16$ rows, $2\cdot10^5$ columns.
The total memory is less than 26MB.
In order to be able to compute hash values for the binning, we need to enclose the data in a $D$-dimensional hypercube, with $M$ linear bins in each dimension. Then the discrete quantized coordinates of the data can be concatenated together to create a feature vector, that can be fed into Count Sketch. 

Question is how to choose the number of bins: 
too many bins will result in very low density in in each bin representing the cluster, to few bins will cause several independent clusters to be merged into one.
Further, we estimate the random collision rate between heavy hitters in adjacent cells.

The discretized volume $V = M^D$ is the total number of bins in the hypercube. We limit the number of heavy hitters as $K < 10^4$.  $\lambda = K/V$ is the mean density of heavy hitters in a cell. We can define a contact neighborhood of the cell by small hypercube with volume $W=3^D$ around it. Given the single cell density $\lambda$, the density of heavy hitters within a  contact neighborhood is
$\rho = W \lambda = K (W/V)$. Since the heavy hitters (from the random collisions perspective) can be treated as a Poisson point process, we can estimate the probability that a neighborhood volume contains 0 or one heavy hitters:
$P(0) = e^{-\rho}, P(>0)=1-P(0)=1-e^{-\rho}$.
The number of heavy hitters with a random collision in its neighborhood is $C = K P(>0)$. This number is quite sensitive on the number of dimensions and the number of linear bins. For $K=10^4$, $D=10$, $M=8$, the collision rate is high: $C=1,057$; while if we increase the number of bins to $M=16$, it goes down to $C=0.00144$. Though only approximately, this argument gives some guidance in choosing the binning. The number of dimensions is limited by the hash collision rate in the sketch matrix, nevertheless, the growth of storage there is only logarithmic (see next section).
It is expected that the hash collisions in the sketch table will cause uncertainties in the estimation of the cell frequency counts. The use of a $\pm1$ hash value for the increment is mitigating this, but still the many cells with small counts will add a fixed Poisson noise to the sketch counts, leading to an increasing relative uncertainty as the frequencies decrease. 
We  evaluated how well CS algorithm estimates the exact frequencies of the discretized multidimensional cells, and how well it ranks those that appear most frequently.  We used the Cancer sample (using 22 bins in each coordinate, top 20K HH) and determined how the relative error grows with the rank of the densest cells. The rank represents a descending ordering, so cells with the highest counts have the lowest ranks. For each cell $i$ we find its frequency $f_i$ and rank $r_i$ in the output of an exact algorithm and its frequency $\hat{f_i}$ in the Count Sketch output. The relative error is defined as $|f_i-\hat{f_i}|/f_i$. The rms values of the relative error are $0.001$ for $r<3000$,  $0.003$ for $3000<r<10000$, and  $0.01$ for $10k<r<20k$.

%% file: 3A-cancer.tex
\section{Applications}
\label{sec:app}
In this paper we test the ideas on two different data sets: (1) 52M pixels in 40 multispectral images from cancer immunotherapy and (2) photometric observations of 30M stars in the Sloan Digital Sky Survey (SDSS).
In both cases the clusters are not thought to be sharply divided into very distinct categories, rather they form distributions where the categories gradually morph into one another. In addition the first few components of PCA do not give a meaningful separation, i.e. nonlinear techniques are needed. However, the data set is too large to use tSNE/UMAP directly. Check project repository \cite{Viska2020} for additional details.

\subsubsection{\textbf{Clustering of pixels in cancer images}}
\label{sec:cancer}
Our dataset consists of 40 images taken at 0.5 $\mu$/pixel resolution, with a 20\% overlap. The slide contains a 5$\mu$ thick section of a melanoma biopsy. The images are observed with a combination of 5 different broadband excitation filters and 20nm wide narrow-band filters, for a total of 35 layers, 1344x1004 pixels in each.  The tissue was stained with 7 different fluorescent markers/dyes: DNA content of nuclei; lineage markers Tumor, CD8, FoxP3 and CD163 (type of cell); and expression markers PD-1 and PD-L1 (``checkpoint blockers'', controlling the interaction between tumors and the immune system).

Cancer cells are mostly located in dense areas, tumors, with a reasonably sharp boundary. Today these tissue areas are annotated visually, by a trained pathologist. Cancer immunotherapy is aimed at understanding the interactions between cancer cells and the immune system, and much of these take place in the Tumor Micro Environment (TME), in the boundary of tumor and immune cells. 

To automate imaging efforts to thousands of images and billions of cells, the task goes beyond detecting and segmenting into distinct cells of a certain type, it is also important to automatically identify the tumor tissue and the tumor~membrane. 

We expect the data to have approximately 8 degrees of freedom.
\begin{figure}[t]
\centering
\includegraphics[width=0.7\linewidth]{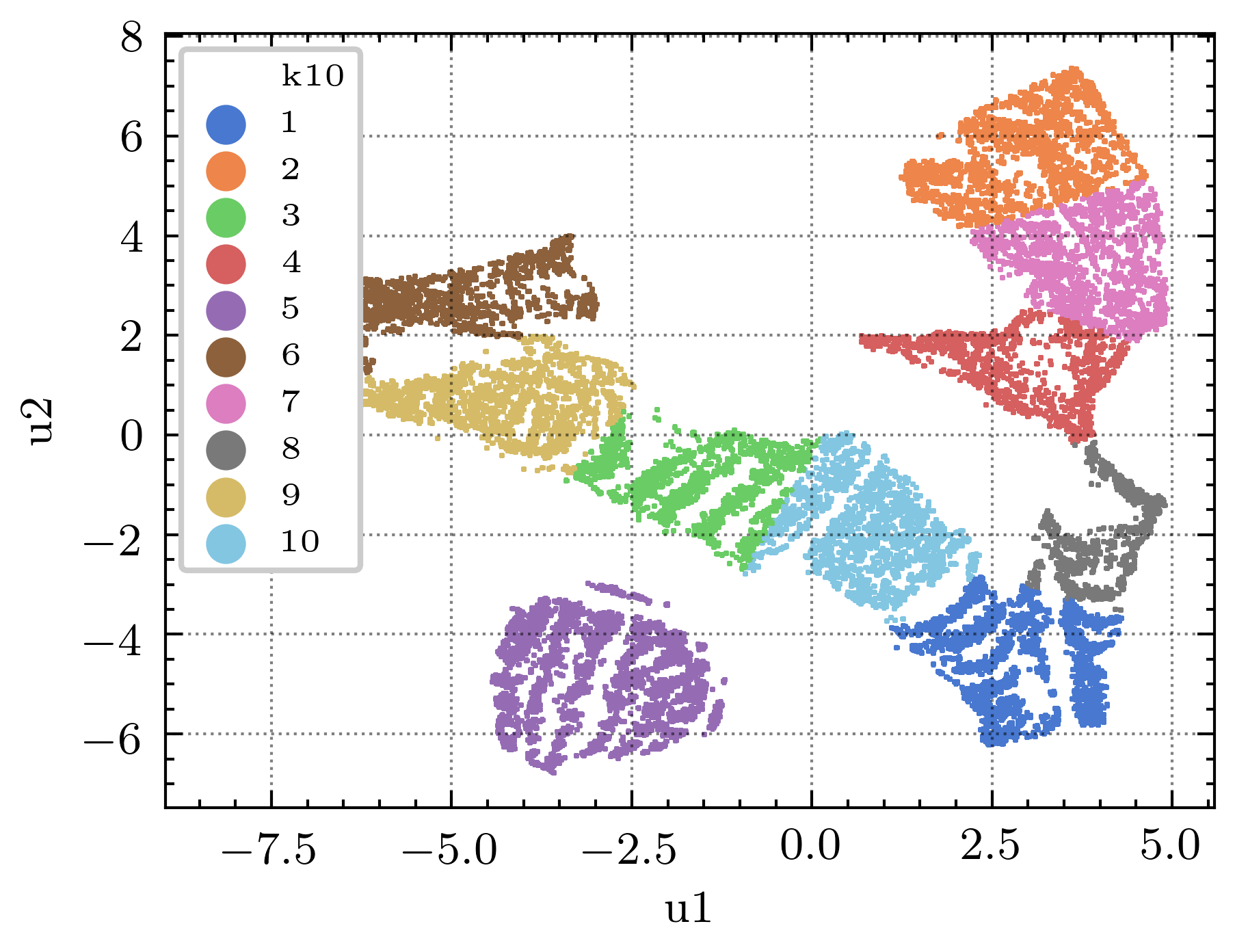}
\caption{\footnotesize The 10 clusters identified in the UMAP processing of the 20,000 top heavy hitters in the cancer data.}
\vspace{-0.6cm}
\label{fig:cancer_umap}
\end{figure}
Our goal is to see what level of clustering can be detected at the pixel level, and whether clustering can be used to identify (a) cells of different types (b) outlines associated with tumor and possibly other tissue types. As each specimen contains different ratios of tissues and cells, naive PCA will not work well, as the weights of the different components will vary from sample to sample. Even if the subspaces will largely overlap, the orientation of the axes will vary from sample to sample. In addition, each staining batch of the samples is slightly different, thus the clusters will be moving around in the pixel color space. Our goal is to verify if we can find enough clusters that can be used further downstream as anchor points for mapping the color space between different staining batches and tissue types. The dataset has a limited set of labels available: a semi-automated segmentation of the images, the detection of cell nuclei and separating them into two basic subtypes, cancer and non-cancer.  The non-cancer cells are likely to have a large fraction of immune cells, but not exclusively. The expression markers are typically attached to the membranes, in between the nuclei.

We take the the first 8 components of the pixel-wise PCA of the images.  We then compute the intensity (Euclidean norm) of the pixel intensities, eliminate the background noise using a threshold derived from the noise. This leaves 26M of the initial 52M pixels. Each pixel is then normalized by the intensity, turning them into colors. 
We embed the points into an 8-dimensional hypercube, and quantize each coordinate into 25 linear bins. Then we run the Count Sketch algorithm and create an ordered list of the top 20,000 heavy hitters. The top HH has 204,901 points, while the 20,000$^{th}$ rank has  only 180.  The cumulative fraction of the top 20,000 heavy hitters is 84.11\%.  The sketch matrix is 16x200,000. We then feed the top 20K HH to UMAP, and generate the top two coordinates. We find 10 clusters in the data, labeled from 1 through 10, as shown on Figure \ref{fig:cancer_umap}. These can be grouped into three categories, pixels related to tumor (1,5,8,10), pixel related to nuclei of cells (6,9), and non-tumor tissue (2,3,4,7). These show an excellent agreement with labels generated for nuclei using an industry standard segmentation software.

\begin{figure}[t]
    \centering
    \includegraphics[width=0.7\linewidth]{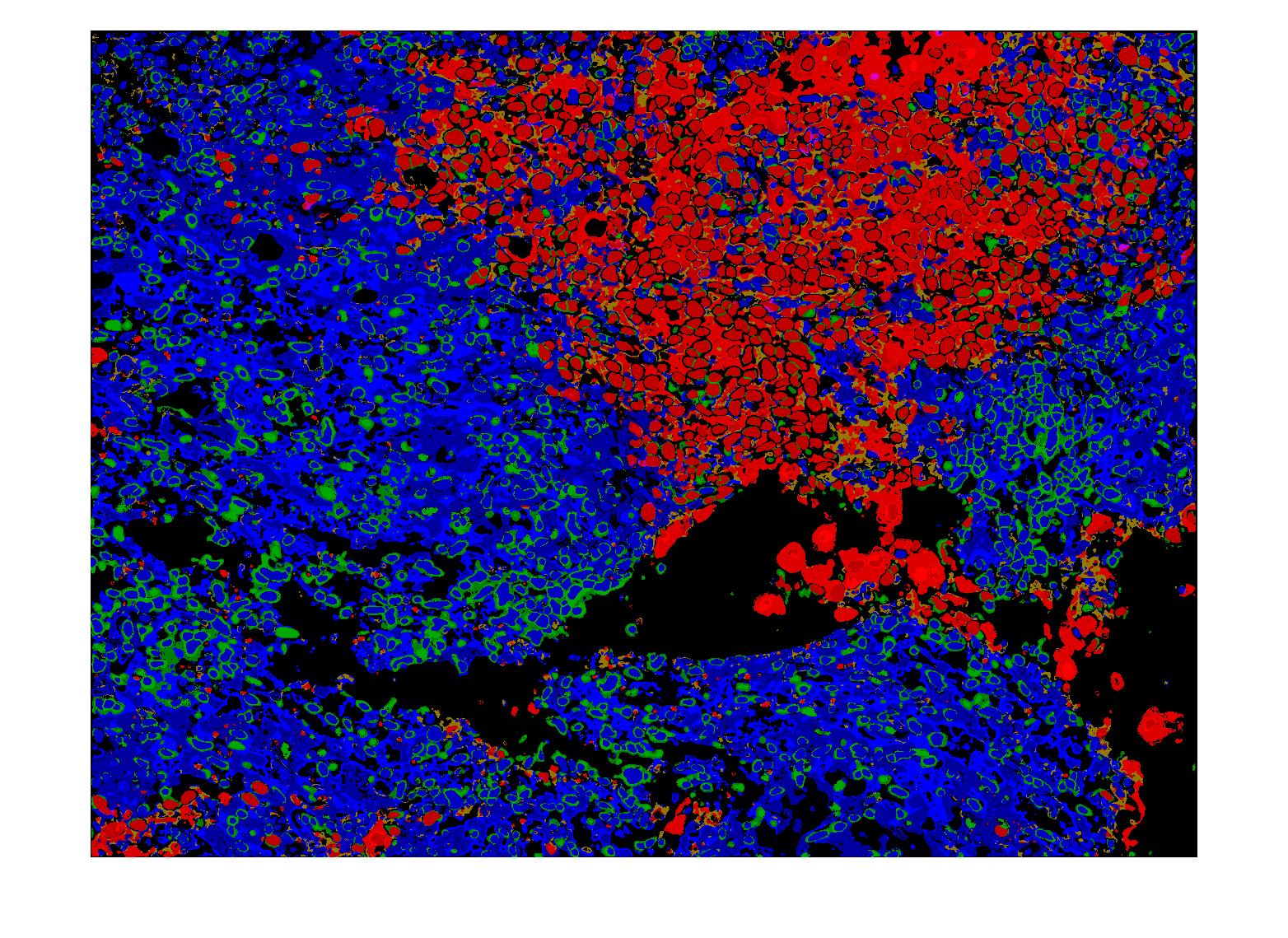}
    \caption{\footnotesize The merged groups in image space: tumor (red), other (blue). The nuclei were painted to the surrounding tissue color.}
    \vspace{-0.6cm}
	\label{fig:cancer_labels}
\end{figure}

\def\arraystretch{0.3}
\begin{table}[ht]
    \centering
    \begin{tabular}{l r r}
        \toprule
    \scriptsize      &\scriptsize  Tumor &\scriptsize  Other  \\
        \midrule
    \scriptsize Tumor &\scriptsize   95792 & \scriptsize  3675   \\ 
    \scriptsize Other &   \scriptsize  6795& \scriptsize  108630 \\
    \bottomrule
    \end{tabular}
    \vspace{-0.2cm}
    \label{tab:classes}
\end{table}
\normalsize
We built a contingency table summarizing the pixel level classifications shown above. For each pixel in the label set marked as Tumor or Other we build a histogram of the 10 classes in our classification. We considered a classification correct if a Tumor pixel in the label set belonged to either a Nucleus or Tumor class in our scheme. We did the same for the Non-Tumor cells.  The pixels tagged as background by our mask were ignored. The results are quite good, the false positive rates are 3.7\% and 5.9\% for Tumor and Other, respectively.

%% file: 3B-sdss.tex
\subsubsection{\textbf{Classification of stellar photometry in SDSS}} \label{sec:stars}

We used the Thirteenth Data Release (DR13) of SDSS. Our goal is to see how well can we recover the traditional astronomical classification of stars, the so called Hertzsprung-Russel diagram. We extracted two different subsets of stars: (1) 540k stars with classification labels obtained from analyzing their spectra and (2) 30M stars without labels. 
The features are the combinations of 5 magnitudes $u,g,r,i,z$, defined by the differences $(ug,ur,...,iz)=(u'-g',u'-r',...,i'-z')$.

\begin{figure}[ht]
    \centering
     \includegraphics[width=0.75\linewidth]{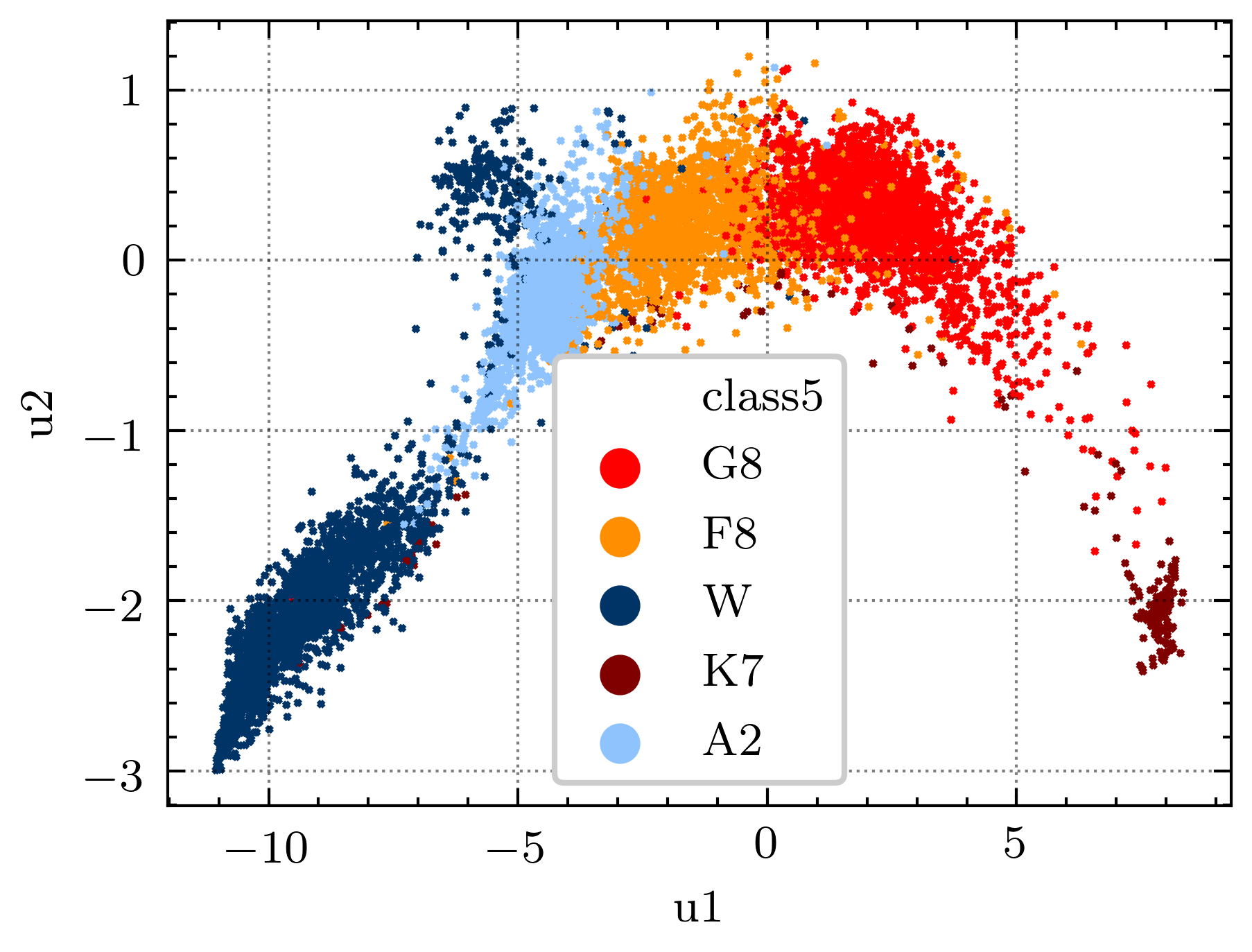}
    \caption{\footnotesize The 2D projection for the u1-u2 UMAP coordinates for stars from SDSS. The UMAP transformation was derived from the 30M photometric sample. The objects shown here are from the five main spectroscopic classes, A2, F8, G8, K7, W. The classes A,F,G,K are ordered by decreasing temperature.}
    \vspace{-0.3cm}
    \label{fig:umap_u1u2}
\end{figure}
We built the Count Sketch on 30M objects and selected the top 2,609 heavy hitters. The count was 1,352,580 at rank 1 and 117 at rank 2,609. The 2,609 HHs contained 99.0\% stars, forming a highly representative sample. We then feed the HHs to UMAP, and extract a 4-dimensional projection. One of the 4-way scatter plot between the coordinates is shown on Figure \ref{fig:umap_u1u2}.
We can distinguish White Dwarfs, and F,A,K and M stars. This experiment has yielded a success beyond any expectations.

%% file: 4-results.tex
\section{Conclusion}

We presented a preprocessing technique that is aimed at reducing the cardinality of extreme sized data sets with moderate dimensions while preserving the clustering properties. We use approximate sketching with a linear time streaming algorithm to find the heavy hitter cells of the quantized input data. This new point process, formed by the heavy hitters will correctly represent the clustering properties of the underlying point cloud. Our code makes heavy use of GPUs for the hash computations and the sketch aggregation, and can be parallelized to an arbitrary high degree. We demonstrated the utility of this approach on two different data sets, one on 50M pixels of cancer images, the other on 30 million stars with 5 colors from the Sloan Digital Sky Survey. We have found that the heavy hitters correctly sampled the clusters in both data sets with quite different properties, and the results were in excellent agreement with the sparse labels available.  

\begin{figure}[t]
    \centering
    \includegraphics[width=0.75
    \linewidth]{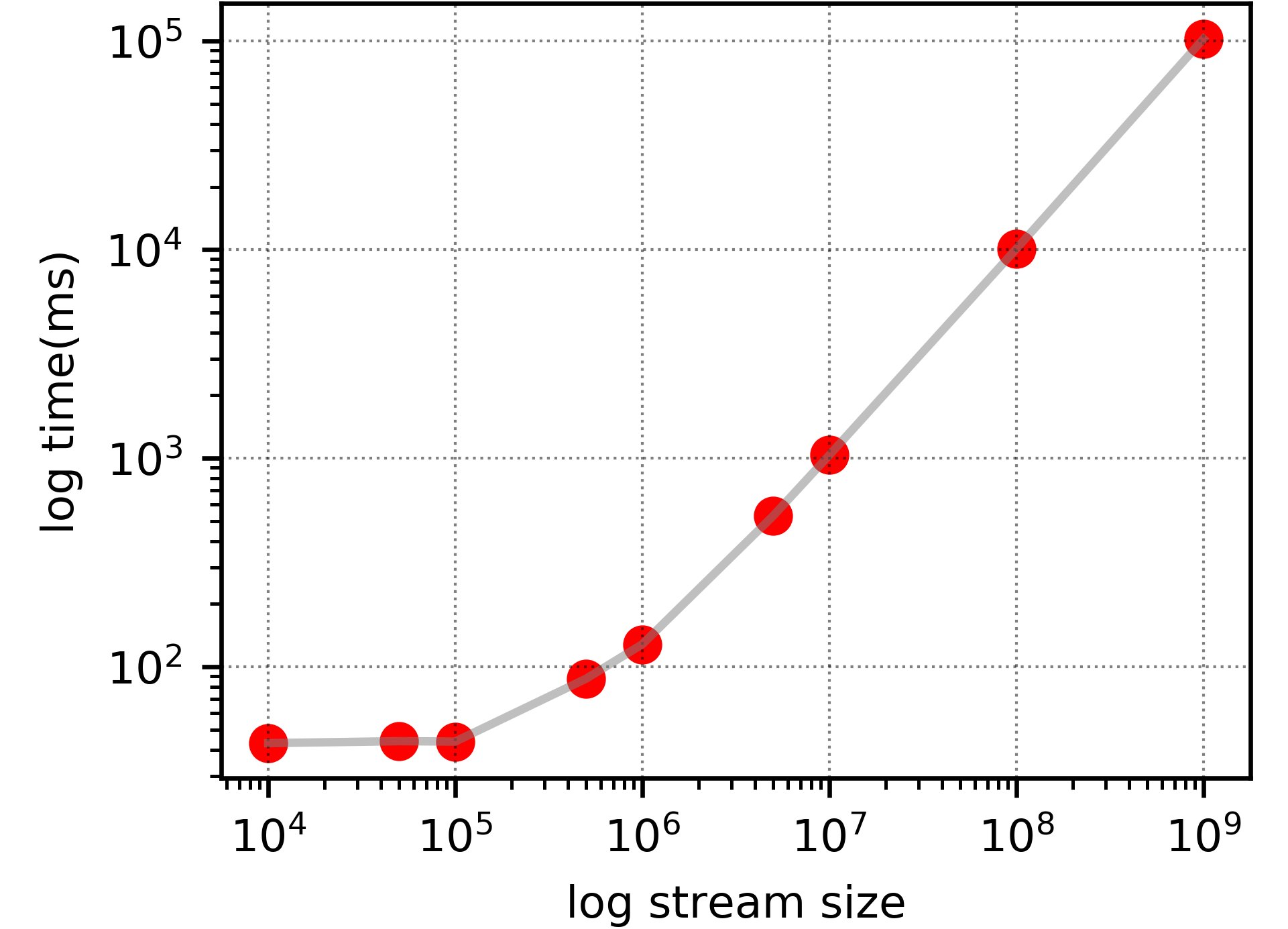}
    \caption{\footnotesize The scaling of the run time for sketch table of size 10x20,000 vs. the size of the data, up to a billion points. For the higher cardinalities the scaling is very linear.}
    \vspace{-0.3cm}
    \label{fig:scaleTime}
\end{figure}

The computations were extremely fast, essentially I/O limited. Processing the Count Sketch of 50 million points takes only a few seconds on a single V100 GPU with a single stream I/O. Introducing parallell I/O would saturate all CUDA kernels of the GPU. By running replicating the data we scaled beyond 1 billion points, and demonstrated an asymptotically linear scaling (Fig. \ref{fig:scaleTime}).
For the cancer data, using pixel colors only we were able to split the pixels into three distinct groups, which have formed spatially coherent and connected regions, separated by the tumor boundaries. This experiment has exceeded our early expectations. The current data used was rather modest, with 52 million pixels in 40 images. Currently at Johns Hopkins University we have more than 45,000 images created, with 100,000 more in the queue, resulting in hundreds of billions of total pixels, as shown on Fig. \ref{fig:scaleTime}. Approaching a pixel-wide analysis of such a data will only be possible through highly scalable algorithms.

For the 30M stars, we generated 20,000 heavy hitters in a very short time. Feeding these to UMAP and projecting to 4 dimensions, let us identify several major classes of stars based upon imaging data only. While it does not represent a breakthrough in astronomy (to properly classify stars we need absolute luminosities, thus distances obtainable only by other techniques) it is a a good demonstration of the scalability and feasibility of our technique.

Our approach has additional long-term implications. Sketches can be computed on arbitrary subsets of the data, and be combined subsequently. The only constraint is that the hashing functions and the sketch matrix sizes must be the same for all threads. Using this approach, sketches of data at different geographic locations can be computed in place, and only the accumulations move to the final aggregation site. This not only saves huge amounts of data movement, but also diminishes potential data privacy concerns, as the approximate hashing is not invertible, i.e. hides all identifiable information. 

Our approach naturally overcomes the problem rising from institutional and national policies limiting the free movement of the data across boundaries and between research centers and hospitals.  Such concerns for studies of  clustering in segregated large-scale data are already present in current Covid-19 research, e.g. aggregating mobility data between cellular providers in different countries. 

When data is stored on a massively parallel storage system, like in many commercial clouds, it is quite easy to run a Count Sketch job over 10000 parallel processes. The sketches can then be aggregated using a tree topology in logarithmic time first within one datacenter, then across many datacenters 
with minimal communications overhead. 


In summary, our algorithm enables the generation of extremely powerful approximate statistics over almost arbitrary large data sets.
